\documentclass[12pt]{article}
\usepackage{amsmath}
\usepackage{amssymb}

\input{tcilatex}
\begin{document}

\bigskip \ 

\bigskip \ 

\begin{center}
\textbf{DIVISION-ALGEBRAS/POINCARE-COJECTURE }

\textbf{\smallskip \ }

\textbf{CORRESPONDENCE}

\smallskip \ 

J. A. Nieto\footnote{%
nieto@uas.edu.mx, janieto1@asu.edu}

\smallskip

\textit{Facultad de Ciencias F\'{\i}sico-Matem\'{a}ticas de la Universidad
Aut\'{o}noma de Sinaloa, 80010, Culiac\'{a}n Sinaloa, M\'{e}xico}

\smallskip \ 

\textit{Mathematical, Computational \& Modeling Sciences Center, Arizona
State University, PO Box 871904, Tempe, AZ 85287, USA}

\bigskip \ 

Abstract

\smallskip \ 
\end{center}

We briefly describe the importance of division algebras and Poincar\'{e}
conjecture in both mathematical and physical scenarios. Mathematically, we
argue that using the torsion concept one can combine the formalisms of
division algebras and Poincar\'{e} conjecture. Physically, we show that both
formalisms may be the underlying mathematical tools in special relativity
and cosmology. Moreover, we explore the possibility that by using the
concept of $n$-qubit system, such conjecture may allow generalization the
Hopf maps.

\bigskip \ 

\bigskip \ 

\bigskip \ 

\bigskip \ 

\bigskip \ 

\bigskip \ 

\bigskip \ 


March, 2013

\newpage

It is known that if there exist a real division algebra then the $n$%
-dimensional sphere $S^{n}$ in $R^{n+1}$ is parallelizable [1]-[3]. It is
also known that the only parallelizable spheres are $S^{1},S^{3}$ and $S^{7}$
[4] (see also Ref. [5]). So one concludes that division algebras only exist
in $1,2,3$ or $8$ dimensions (see Refs. [6]-[10] and references therein). It
turns out that these theorems are deeply related to the Hopf maps, $S^{3}%
\overset{S^{1}}{\longrightarrow }S^{2}$, $S^{7}\overset{S^{3}}{%
\longrightarrow }S^{4}$ and $S^{15}\overset{S^{7}}{\longrightarrow }S^{8}$
[4]. Focusing on $S^{3}$, it is intriguing that none of these remarkable
results seem to have been considered in the proof the the original Poincar%
\'{e} conjecture [11]-[13], which establishes that any closed simply
connected $3$-manifold $\mathcal{M}^{3}$ is homeomorphic to $S^{3}$. In
fact, until now any proof of the Poincar\'{e} conjecture associated with $%
S^{3}$ is based in the Ricci flow equation [14] (see also Refs. [11]-[13]),
but the parallelizabilty of $S^{3}$ (or any $M^{3}$ manifold) is not even
mentioned. The main goal of this work is to establishes a link between the
concept of parallelizability and the Ricci flow equation. We also explain a
number of physical scenarios where such a link may be important, including
special relativity, cosmology and Hopf maps via $n$-qubit systems (see Ref.
[15] and Refs. therein).

Before we address the problem at hand it is worth making a number of
comments. Let us start mentioning that it has been shown that division
algebras are linked to different physical scenarios, including, superstrings
[16] and supersymmetry [17]-[18]. Even more surprising is the fact that
division algebras are also linked to quantum information theory via the $n$%
-qubit theory (see Refs. [19]-[21]). Mathematically, division algebras are
also connected with important arenas such as K-theory [6]. If a division
algebra is normed then one may also introduce the four algebras; real
numbers, complex numbers, quaternions and octonions (see Ref. [10]). On the
other hand the Poincar\'{e} conjecture seems to be useful in the discussion
of various cosmological models (see Refs. [22]-[25]) and the study of
gravitational instanton theory [26].

One may ask ourselves: Are all this links a coincidence? or there is in
these links a deep underlying message? An indicator that starting with
division algebras one may obtain a deep physical result is illustrated by
superstrings. In fact, in this case the dimensionality of the spacetime it
is not putted by hand but is a prediction of the theory. It turns out that
at the quantum level one finds a consistent superstring theory only when the
dimension of the spacetime $D$ takes values in the set $E=\{3,4,6,10\}$.
Considering light-like coordinates such that $n=D-2$ one realizes that $E$
can be reduced to the set $\mathcal{E}=\{1,2,4,8\}$. But $\mathcal{E}$
corresponds exactly to the only dimensions where a division algebra may
exist (see Ref. [16] for details). From this perspective one may say that in
a sense the dimensions where a quantum consistent superstring theory may
exist are predicted by division algebras. Another scenario where the
division-algebra/Poincar\'{e}-conjecture correspondence may play a physical
important role is in instanton theory. In this case the Hopf maps determine
the different structures of instanton solutions (see Refs [26]).

Let us start introducing the metric tensor

\begin{equation}
\gamma _{ab}=\gamma _{ab}(x^{c}).  \tag{1}
\end{equation}%
Here, $x^{a}$ is a coordinate patch in a $n$-dimensional manifold $M^{n}$.
We also introduce a Riemann symmetric connection $\Gamma _{ab}^{c}=\Gamma
_{ba}^{c}$ and the totally antisymmetric torsion tensor $%
T_{ab}^{c}=-T_{ba}^{c}$. Geometric parallelizability of $M^{n}$ means the
\textquotedblleft flattening\textquotedblright \ the space in the sense that

\begin{equation}
\mathcal{R}_{bcd}^{a}(\Omega _{fg}^{e})=0,  \tag{2}
\end{equation}%
where

\begin{equation}
\mathcal{R}_{bcd}^{a}=\partial _{c}\Omega _{bd}^{a}-\partial _{d}\Omega
_{bc}^{a}+\Omega _{ec}^{a}\Omega _{bd}^{e}-\Omega _{ed}^{a}\Omega _{bc}^{e},
\tag{3}
\end{equation}%
is the Riemann curvature tensor, with

\begin{equation}
\Omega _{ab}^{c}=\Gamma _{ab}^{c}+T_{ab}^{c}.  \tag{4}
\end{equation}%
By substituting (4) into (3) one finds

\begin{equation}
R_{bcd}^{a}+D_{c}T_{bd}^{a}-D_{d}T_{bc}^{a}+T_{ec}^{a}T_{bd}^{e}-T_{ed}^{a}T_{bc}^{e}=0.
\tag{5}
\end{equation}%
Here, $D_{c}$ denotes a covariant derivative with $\Gamma _{ab}^{c}$ as a
connection and

\begin{equation}
R_{bcd}^{a}=\partial _{c}\Gamma _{bd}^{a}-\partial _{d}\Gamma
_{bc}^{a}+\Gamma _{ec}^{a}\Gamma _{bd}^{e}-\Gamma _{ed}^{a}\Gamma _{bc}^{e}.
\tag{6}
\end{equation}%
Using in (5) the cyclic identities for $R_{bcd}^{a}$ one gets

\begin{equation}
D_{c}T_{bda}=T_{e[bd}T_{a]c}^{e},  \tag{7}
\end{equation}%
where

\begin{equation}
T_{e[bd}T_{a]c}^{e}\equiv \frac{1}{3}%
\{T_{ebd}T_{ac}^{e}+T_{eab}T_{dc}^{e}+T_{eda}T_{bc}^{e}\}.  \tag{8}
\end{equation}%
Substituting (7) into (5) one obtains the key formula

\begin{equation}
R_{abcd}^{\ }=T_{eab}T_{cd}^{e}-T_{e[ab}T_{c]d}^{e}.  \tag{9}
\end{equation}

For a $n$-dimensional sphere $S^{n}$ with radius $l$ we have, $\gamma
_{ab}\longrightarrow g_{ab}$,

\begin{equation}
R_{abcd}^{\ }=\frac{1}{l^{2}}(g_{ac}g_{bd}-g_{ad}g_{bc}),  \tag{10}
\end{equation}%
where $g_{ab}$ is the metric on $S^{n}$, and therefore one gets the
expression

\begin{equation}
\frac{1}{l^{2}}%
(g_{ac}g_{bd}-g_{ad}g_{bc})=T_{eab}T_{cd}^{e}-T_{e[ab}T_{c]d}^{e}.  \tag{11}
\end{equation}%
Contracting in (11) with $g^{ac}$ and $T_{f}^{ac}$ it leads to first and the
second Cartan-Shouten equations%
\begin{equation}
T_{a}^{cd}T_{bcd}=(d-1)l^{-2}g_{ab},  \tag{12}
\end{equation}%
and

\begin{equation}
T_{ea}^{d}T_{db}^{f}T_{fc}^{e}=(d-4)l^{-2}T_{abc},  \tag{13}
\end{equation}%
respectively. Durander, Gursey and Tze [27] noted that (12) and (13) are
mere covariant forms of the algebraic identities derived in normed division
algebras. It turns out that (12) and (13) can be used eventually to prove
that the only parallelizable spheres are $S^{1},S^{3}$ and $S^{7}$ [5]. In
general, however, for other $n$-manifolds $M^{n}$ the expressions (11)-(13)
does not hold.

If the only condition is that $M^{n}$ is parallelizable one may start with
(9) instead of (11). In this case one finds that contracting (9) with $%
g^{ac} $ leads to

\begin{equation}
R_{ab}=T_{a}^{cd}T_{bcd}.  \tag{14}
\end{equation}%
Here, $R_{ab}=R_{acb}^{c}$ is the Ricci tensor.

Before we relate (14) with de Ricci flow equation used in the Poincar\'{e}
conjecture let us recall how (10) is obtained. We shall focus in $S^{3}$,
but in straightforward way one can generalize the method to any $n$-sphere.
Consider the line element

\begin{equation}
ds^{2}=dx^{2}+dy^{2}+dz^{2}+dw^{2}.  \tag{15}
\end{equation}%
The sphere $S^{3}$ can be defined by the constraint

\begin{equation}
x^{2}+y^{2}+z^{2}+w^{2}=l^{2},  \tag{16}
\end{equation}%
where $l$ is constant. From (16) one sees that

\begin{equation}
w=(l^{2}-(x^{2}+y^{2}+z^{2}))^{1/2}.  \tag{17}
\end{equation}%
Rigorously, one must write $w=\epsilon (l^{2}-(x^{2}+y^{2}+z^{2}))^{1/2}$,
with $\epsilon =\pm 1$. But it turns out that our computations are
independent of $\epsilon $. Furthermore, it will be useful for further
computations to write (15) and (17) in the form

\begin{equation}
ds^{2}=dx^{a}dx^{b}\delta _{ab}+dw^{2}  \tag{18}
\end{equation}%
and

\begin{equation}
w=(l^{2}-x^{a}x^{b}\delta _{ab})^{1/2},  \tag{19}
\end{equation}%
respectively. The symbol $\delta _{ab}$ is a Kronecker delta. From (19) one
obtains%
\begin{equation}
dw=\frac{-x^{a}dx_{a}}{(l^{2}-x^{c}x^{d}\delta _{cd})^{1/2}},  \tag{20}
\end{equation}%
where $x_{a}=x^{b}\delta _{ab}$. So, substituting (20) into (18) yields the
line element

\begin{equation}
ds^{2}=dx^{a}dx^{b}g_{ab},  \tag{21}
\end{equation}%
with

\begin{equation}
g_{ab}=\delta _{ab}+\frac{x_{a}x_{b}}{(l^{2}-x^{c}x^{d}\delta _{cd})}. 
\tag{22}
\end{equation}%
The inverse $g^{ab}$ of $g_{ab}$ is given by%
\begin{equation}
g^{ab}=\delta ^{ab}-\frac{x^{a}x^{b}}{l^{2}}.  \tag{23}
\end{equation}%
Moreover, using (22) and (23) one finds that the Christoffel symbols $\Gamma
_{cd}^{a}$ become

\begin{equation}
\Gamma _{cd}^{a}=\frac{1}{l^{2}}x^{a}g_{cd}.  \tag{24}
\end{equation}%
Considering (6), it is straightforward to see that the Riemann curvature
tensor associated with (24) is given by the expression (10).

Now we would like to generalize the key constraint (19) in form

\begin{equation}
w=\varphi (x^{a}),  \tag{25}
\end{equation}%
where $\varphi $ is an arbitrary function of the coordinates $x^{a}$. In
this case, the metric $\gamma _{ab}$ becomes

\begin{equation}
\gamma _{ab}=\delta _{ab}+\partial _{a}\varphi \partial _{b}\varphi , 
\tag{26}
\end{equation}%
while the inverse $\gamma ^{ab}$ is given by

\begin{equation}
\gamma ^{ab}=\delta ^{ab}-\frac{\partial ^{a}\varphi \partial ^{b}\varphi }{%
1+\partial ^{c}\varphi \partial _{c}\varphi }.  \tag{27}
\end{equation}%
The Christoffel symbols become%
\begin{equation}
\Gamma _{cd}^{a}=\frac{\partial ^{a}\varphi \partial _{cd}\varphi }{%
1+\partial ^{e}\varphi \partial _{e}\varphi }.  \tag{28}
\end{equation}%
After lengthy but straightforward computation one discovers that the Riemann
tensor $R_{abcd}^{\ }$ obtained form (28) is%
\begin{equation}
R_{abcd}=\frac{1}{1+\partial ^{e}\varphi \partial _{e}\varphi }(\partial
_{ac}\varphi \partial _{bd}\varphi -\partial _{ad}\varphi \partial
_{bc}\varphi ).  \tag{29}
\end{equation}%
One can verifies that when one considers the particular case

\begin{equation}
\varphi =(l^{2}-x^{a}x^{b}\delta _{ab})^{1/2},  \tag{30}
\end{equation}%
then (10) follows from (29).

Let us now consider the Ricci flow evolution equation [14] (see also Refs.
[11]-[13] and references therein)

\begin{equation}
\frac{\partial \gamma _{ab}}{\partial t}=-2R_{ab}.  \tag{31}
\end{equation}%
Here, as before, $R_{ab}=R_{acb}^{c}$ is the Ricci tensor. In this case the
metric $\gamma _{ab}(t)$ is understood as a family of Riemann metrics on $%
M^{3}$. It has been emphasized that the Ricci flow equation is the analogue
of the heat equation for metrics $\gamma _{ab}$. The central idea is that a
metric $\gamma _{ab}$ associated with a closed simply connected manifold $%
\mathcal{M}^{3}$ evolves according to (31) towards a metric $g_{ab}$ of $%
S^{3}$. Symbolically, this means that in virtue of (31) we have the metric
evolution $\gamma _{ab}\longrightarrow g_{ab}$, which in turn must imply the
homeomorphism $\mathcal{M}^{3}\longrightarrow S^{3}$.

The question arises whether one can introduce the parallelizability concept
into (31). Let us assume that $\mathcal{M}^{3}$ is a parallelizable manifol.
We shall also assume that $\mathcal{M}^{3}$ is determined by the general
constraint (25). First observe that using (14), in this case the Ricci
equation (31) can be written as

\begin{equation}
\frac{\partial \gamma _{ab}}{\partial t}=-2T_{a}^{cd}T_{bcd}.  \tag{32}
\end{equation}%
This is interesting result because it means that the evolution of $\gamma
_{ab}$ is determined by the torsion tensor $T_{bc}^{a}$. Moreover, combining
(9) and (29) one derives the formula

\begin{equation}
\frac{1}{1+\partial ^{e}\varphi \partial _{e}\varphi }(\partial _{ac}\varphi
\partial _{bd}\varphi -\partial _{ad}\varphi \partial _{bc}\varphi
)=T_{eab}T_{cd}^{e}-T_{e[ab}T_{c]d}^{e},  \tag{33}
\end{equation}%
which, using (26), allows to write (32) in the form

\begin{equation}
\partial _{a}\dot{\varphi}\partial _{b}\varphi +\partial _{a}\varphi
\partial _{b}\dot{\varphi}=-\frac{2\gamma ^{cd}}{1+\partial ^{e}\varphi
\partial _{e}\varphi }(\partial _{ac}\varphi \partial _{bd}\varphi -\partial
_{ad}\varphi \partial _{bc}\varphi ).  \tag{34}
\end{equation}%
In the case of $S^{3}$ manifold, using (10) or (12) one obtains a Einstein
type metric

\begin{equation}
R_{ab}=\frac{2}{l^{2}}g_{ab}  \tag{35}
\end{equation}%
and the evolution equation becomes

\begin{equation}
\frac{\partial g_{ab}}{\partial t}=-\frac{4}{l^{2}}g_{ab}.  \tag{36}
\end{equation}%
This type of equation are discussed extensively in references [11] and [13].
The relevant features is that from the solution one sees that at large times
evolution behavior of $g_{ab}$ is $g_{ab}(t)=(1-\frac{2}{l^{2}}t)g_{ab}(0)$,
where $g_{ab}(0)$ corresponds to an initial condition for the metric. In
this case one has $R_{ab}(t)=R_{ab}(0)$ and therefore since $\frac{2}{l^{2}}%
>0$ one has uniform contraction with singularity at $t=\frac{l^{2}}{2}$ (see
Ref. [13] for details).

Let us now discuss some physical scenarios where the division-algebra/Poincar%
\'{e}-conjecture correspondence may be relevant. Let us start by first
recalling the Einstein field equations with cosmological constant $\Lambda $,

\begin{equation}
R_{ab}-\frac{1}{2}\gamma _{ab}R+\Lambda \gamma _{ab}=0.  \tag{37}
\end{equation}%
It is known that the lowest energy solution of (37) corresponds precisely to 
$S^{3}$ (or to $S^{n}$ in general). In this case the cosmological constant $%
\Lambda $ is given by $\Lambda =\frac{2}{l^{2}}$. This can be verified using
(10) and (37) (Actually this solution can be understood as a De Sitter type
solution.) The question arises: how can be understood a metric solution $%
\gamma _{ab}$ of (37) associated with both $\mathcal{M}^{3}$ and the Ricci
flow equation? Thinking about quantum mechanics analogue one may argue that
one may visualize $\mathcal{M}^{3}$ as a excite state which, according to
the Poincar\'{e} conjecture, must decay (homeomorphically) to $S^{3}$.
Symbolically one may write this as $\mathcal{M}^{3}\rightarrow S^{3}$.

Considering the transition $\mathcal{M}^{3}\rightarrow S^{3}$ we discover
that even in special relativity one may find this kind process. Consider the
well-known time dilatation formula

\begin{equation}
dt=\frac{d\tau }{\sqrt{1-\frac{\mathbf{v}^{2}}{c^{2}}}}.  \tag{38}
\end{equation}%
Here, of course $\tau $ is the proper time, $c$ is the light velocity and we
are thinking $\mathbf{v}$ as the velocity of the relativistic object in
three dimensions, namely $\mathbf{v}^{2}=v_{x}^{2}+v_{y}^{2}+v_{z}^{2}$. It
is not difficult to see that (38) can also be written as

\begin{equation}
v_{x}^{2}+v_{y}^{2}+v_{z}^{2}+v_{0}^{2}=c^{2},  \tag{39}
\end{equation}%
where $v_{0}=\frac{d(c\tau )}{dt}$. One can understand the constraint (39)
as a formula in the space of velocities (tangent space) which determines a $%
S_{v}^{3}$ manifold. So, one wonders what could be the corresponding
generalized $3$-manifold $\mathcal{M}_{v}^{3}$. One may consider in an
extension of (39) in the form

\begin{equation}
v_{0}=\varphi (v_{x},v_{y},v_{z}).  \tag{40}
\end{equation}%
But in this case, the question arises whether the light velocity $c$ itself
may be understood as excited state $\mathcal{C}$. Hence, the evolution
process $\varphi (v_{x},v_{y},v_{z})\rightarrow \sqrt{%
c^{2}-(v_{x}^{2}+v_{y}^{2}+v_{z}^{2})}$ may be understood as the transition $%
\mathcal{C}\rightarrow c$. This may be relevant to consider the light
velocity $c$ no as a given constant but as a result of evolution transition.
It may be interesting to see what the torsion means in this context.

In a cosmology context we also find a possible application of the
division-algebra/Poincar\'{e}-conjecture link. It is known that
topologically, the standard Friedmann-Lemaitre-Robertson-Walker universe
corresponds to a time evolving radius of a $S^{3}$ space. In reference [22]
it argue that if this universe is modified in $\mathcal{M}^{3}$, at the end
the acceleration may produce a phase transition changing $\mathcal{M}^{3}$
to a space of constant curvature which corresponds precisely de Sitter phase
associated with $S^{3}$. Another point of view is that since the Thurston
three-dimensional geometrization conjecture (a generalization of the Poincar%
\'{e} conjecture) requires one to understand all locally homogeneous
geometries on closed three manifolds, using Ricci flow one may consider
Bianchi classes (see Ref. [25] for details) used to study cosmological
models in a general context [28]. What one may add to this scenario is that
such a transition may require a torsion in order to make $S^{3}$ (or other
Bianchi cosmological models) parallelizable.

We would like also to describe an application of Division-algebra/Poincar%
\'{e}-conjecture correspondence in qubits theory. It has been mentioned in
Ref. [19], and proved in Refs. [20] and [21], that for normalized qubits the
complex $1$-qubit, $2$-qubit and $3$-qubit are deeply related to division
algebras via the Hopf maps, $S^{3}\overset{S^{1}}{\longrightarrow }S^{2}$, $%
S^{7}\overset{S^{3}}{\longrightarrow }S^{4}$ and $S^{15}\overset{S^{7}}{%
\longrightarrow }S^{8}$, respectively. It seems that there does not exist a
Hopf map for higher $N$-qubit states. Therefore, from the perspective of
Hopf maps, and therefore of division algebras, one arrives to the conclusion
that $1$-qubit, $2$-qubit and $3$-qubit are more special than higher
dimensional qubits (see Refs. [19]-[21] for details). Considering the $2$%
-qubit as a guide. One notice that $S^{3}$ plays the role of fiber in the
map $S^{7}\overset{S^{3}}{\longrightarrow }S^{4}$. Thus, in principle one
may think in a more general map $\mathcal{M}^{7}\overset{\mathcal{M}^{3}}{%
\longrightarrow }\mathcal{M}^{4}$ in turn this may lead to a more general $2$%
-qubit system, which one may call $2$-Poinqubit (just to remember that this
is a concept inspired by Poincar\'{e} conjecture.) At the end one may be
able to obtain the transition $2$-Poinqubit$\longrightarrow 2$-qubit. Of
course one may extend most of the arguments develop in this work to the
other Hopf maps $S^{3}\overset{S^{1}}{\longrightarrow }S^{2}$ and $S^{15}%
\overset{S^{7}}{\longrightarrow }S^{8}$.

Finally, it is tempting to speculate about two other topics where our
formalism may have some interest. The first one refers about a possible
generalization of the Ricci flow equation (31) to a complex context. In this
case the metric $\gamma _{ab}$ and the Ricci tensor $R_{ab}$ may be
complexified $\gamma _{ab}\rightarrow \psi _{ab}$ and $R_{ab}\rightarrow 
\mathcal{R}_{ab}$, respectively. But if this is the case then instead of
(31) one must consider a Schr\"{o}dinger type equation

\begin{equation}
i\frac{\partial \psi _{ab}}{\partial t}=-2\mathcal{R}_{ab},  \tag{41}
\end{equation}%
for the evolving complex metric $\mathcal{\gamma }_{ab}$. The second topic
is about a possible connection of the Poincar\'{e} conjecture with oriented
matroid theory [29] (see also Refs. [30]-[35] and references therein). This
is because to any sphere $S^{n}$ one may associate a polyhedron which under
stereographic projection corresponds to a graph in $R^{n+1}$. It turns out
that matroid theory can be understood as a generalization of graph theory
and therefore it may be interesting to see if there is any connection
between oriented matroid theory and Poincar\'{e} conjecture. In fact in
oriented matroid theory there exist the concept of pseudo-spheres which
generalize the ordinary concept of spheres (see Ref. [29] for details). So
one wonders wether there exist the analogue of Poincar\'{e} conjecture for
pseudo-spheres.

\bigskip \ 

\noindent \textbf{Acknowledgments: }I would like to thank the Mathematical,
Computational \& Modeling Science Center of the Arizona State University
where part of this work was developed. This work was partially supported by
Profapi/2012.

\end{document}